\documentclass[12pt]{article}%
\usepackage{amsmath}
\usepackage{amsfonts}
\usepackage{amssymb}
\usepackage{graphicx}%
\providecommand{\U}[1]{\protect\rule{.1in}{.1in}}
\begin{document}
\begin{titlepage}
\ \\
\begin{center}
\LARGE
{\bf
Comment on\\
"When the Quantum Energy Teleportation is Observable? "
}
\end{center}
\ \\
\begin{center}
\large{
Masahiro Hotta
}\\
\ \\
\ \\
{\it
Department of Physics, Faculty of Science, Tohoku University,\\
Sendai, 980-8578, Japan\\
hotta@tuhep.phys.tohoku.ac.jp
}
\end{center}
\begin{abstract}
Recently authors of a paper (arXiv: 2105.04407) claim that quantum energy teleportation is unobservable due to time-energy uncertainty relation.
In this short note, I will point out that their argument is wrong. They misuse the uncertainty relation.
\end{abstract}
\end{titlepage}

\bigskip

Recently Razmi and MohammadKazemi wrote a paper about quantum energy
teleportation (QET) \cite{rm} and claim that QET is unobservable due to
time-energy uncertainty relation. In this short note, I will point out that
their argument is wrong. They use a wrong uncertainty relation.

Before going to clarify the flaw of their argument, let me remind readers of
the minimal QET model \cite{qet} the authors used. The system consists of two
distant qubits: qubit $A$ which Alice possesses and qubit $B$ which Bob
possesses. The Hamiltonian is the same as that of the Ising spin chain in the
presence of transeversal magnetic field as follows:%
\begin{equation}
H=H_{A}+H_{B}+V,\label{0}%
\end{equation}
where each contribution is given by
\begin{align}
H_{A} &  =h\sigma_{A}^{z}+\frac{2h^{2}}{\sqrt{4h^{2}+k^{2}}},\label{1}\\
H_{B} &  =h\sigma_{B}^{z}+\frac{2h^{2}}{\sqrt{4h^{2}+k^{2}}},\\
V &  =k\sigma_{A}^{x}\sigma_{B}^{x}+\frac{k^{2}}{\sqrt{4h^{2}+k^{2}}%
},\label{3}%
\end{align}
and $h$ and$~k$ are positive constants with energy dimension, $\sigma_{A}%
^{x}~\left(  \sigma_{B}^{x}\right)  $ is the x-component of the Pauli
operators for the qubit of A (B) and $\sigma_{A}^{z}~\left(  \sigma_{B}%
^{z}\right)  $ is the z-component for the qubit of A (B). The constant terms
in Eq. (\ref{1})-Eq. (\ref{3}) are added in order to make the expectational
value of each operator zero for the ground state $|g\rangle$:%

\[
\langle g|H_{A}|g\rangle=\langle g|H_{B}|g\rangle=\langle g|V|g\rangle=0.
\]
Because the lowest eigenvalue is $E_{0}=0$, the Hamiltonian $H$ is
non-negative operator. The ground state is given by
\[
|g\rangle=\frac{1}{\sqrt{2}}\sqrt{1-\frac{2h}{\sqrt{4h^{2}+k^{2}}}}%
|+\rangle_{A}|+\rangle_{B}-\frac{1}{\sqrt{2}}\sqrt{1+\frac{2h}{\sqrt
{4h^{2}+k^{2}}}}|-\rangle_{A}|-\rangle_{B},
\]
where $|\pm\rangle_{A}~\left(  |\pm\rangle_{B}\right)  $ are eigenstates of
$\sigma_{A}^{z}~\left(  \sigma_{B}^{z}\right)  $ with eigenvalues $\pm1$.

In the QET protocol, Alice first performs $\sigma_{A}^{x}$ measurement of $A$.
The measurement result is represented as $(-1)^{\mu}$ with $\mu=0,1$. The
projection operators $P_{A}(\mu)$ for the post-measurement state $P_{A}%
(\mu)|g\rangle$ are given by
\[
P_{A}(\mu)=\frac{1}{2}\left(  I+\left(  -1\right)  ^{\mu}\sigma_{A}%
^{x}\right)  .
\]
After the measurement, $A$ is in an excited state and has an expectation value
of energy given by%
\[
E_{A}=\frac{h^{2}}{\sqrt{h^{2}+k^{2}}}.
\]
The energy distribution is localized at $A$ soon after the measurement, but
the dynamics induced by $H$ transfers the energy to $B$ as%
\[
\langle H_{B}(t)\rangle=\frac{h^{2}}{2\sqrt{h^{2}+k^{2}}}\left[  1-\cos\left(
4kt\right)  \right]  .
\]
The time-scale order of energy transfer is provided by $1/k$. Alice informs
the measurement result to Bob in a time duration $t_{teleportation}$ much
shorter than $1/k$:%

\begin{equation}
t_{teleportation}\ll1/k.\label{99}%
\end{equation}

Then by using a quantum operation device $C$ interacting with $B$, Bob
performs unitary operation%

\[
U_{B}(\mu)=I_{B}\cos\theta-i(-1)^{\mu}\sigma_{B}^{y}\sin\theta
\]
to $B$, where the real parameter $\theta$ is fixed so as to move the maximum
energy from $B$ to $C$. The value of the maximum energy is postive and is
given by
\[
E_{B}=\frac{2h^{2}+k^{2}}{\sqrt{4h^{2}+k^{2}}}\left[  \sqrt{1+\frac{h^{2}%
k^{2}}{\left(  2h^{2}+k^{2}\right)  ^{2}}}-1\right]  .
\]
This is the minimal QET.

Razmi and MohammadKazemi \cite{rm} estimate the value of $E_{B}$ as
\begin{equation}
E_{B}\leq0.13k\label{100}%
\end{equation}
and argue that the QET cannot be observable due to an obstacle. They propose a
time-energy uncertainty relation as%
\begin{equation}
\Delta E_{B}\Delta t_{teleportation}\geq1\label{101}%
\end{equation}
and, by using eq.(\ref{101}), conclude that
\begin{equation}
E_{B}t_{teleportation}\geq1.\label{103}%
\end{equation}
Using eq. (\ref{99}) and eq. (\ref{103}), they derive
\begin{equation}
E_{B}\gg k.\label{102}%
\end{equation}
Apparently eq.(\ref{100}) and eq. (\ref{102}) contradict with each other. Thus
they conclude that the mimimal QET is unobservable .

Unfortunately, their argument is wrong because their time-energy uncertainty
relation in eq. (\ref{101}) does not hold. The teleportation time
$t_{teleportation}$ does not have any qunatum fluctuation and can be fixed as
$t_{teleportation}=\epsilon/k$, where $\epsilon$ is a small number just like
$10^{-3}$. The coupling constant $k$ is not a degree of freedom of the system,
and does not change in time. Before the QET experiement, we are able to
measure the value of $k$ in an arbitrary precision. Also we are able to adopt
a classical channel between Alice and Bob, which achieves $t_{teleportation}$,
even though the energy cost $E_{cc}$ of the classical communication via the
channel is high. The teleported energy $E_{B}$ is additional energy gain
independent of $E_{cc}$ for Bob. Since $t_{teleportation}$ is fixed as
$\epsilon/k$ by the coupling constant $k$, $t_{teleportation}$ does not have
any quantum fluctution and measurement error like $\Delta t_{teleportation}$
proposed by them:%
\[
\Delta t_{teleportation}=0.
\]
Also no measurement error $\Delta E_{B}$ of $E_{B}$ appears. The teleported
energy $E_{B}$ is moved to the quantum operation device $C$ of Bob. By taking
arbitrary long measurement time for $C$, Bob is able to measure $E_{B}$ stored
in $C$ in an arbitrary precision. This implies
\[
\Delta E_{B}=0.
\]
Thus the uncertainty relation proposed by them in eq.(\ref{101}) cannot be
satsified.In conclusion, their argument never indicate the impossibility of
observation of the minimal QET.


\begin{thebibliography}{9}                                                                                                %
\bibitem {rm}H. Razmi, A. MohammadKazemi, "When the Quantum Energy
Teleportation is Observable?", arXiv:2105.04407.

\bibitem {qet}M. Hotta. A protocol for quantum energy distribution, Phys.
Lett. A 372, 5671 (2008).
\end{thebibliography}
\end{document}